# Equilibrium and overcritical solitons at a spin-flop phase transition

**V.V.Nietz**

*Joint Institute for Nuclear Research,*
*Dubna, Moscow Region 141980, Russia*
*E-mail*: nietz@jinr.ru, Tel: +79060864067, Fax: +7-496-21-65-882

## Abstract

Taking into account the energy dissipation in the spin-flop phase transition induced by magnetic field in an antiferromagnet with uni-axial anisotropy, the following peculiarities of precessing ball solitons (PBS) are considered: a) the states of equilibrium solitons arising in the initial phase during the phase transition; b) the states of "overcritical PBS", existing outside the region of metastability. The "overcritical" PBS may originate during disintegration of the initial phase.

*PACS*: 75.25; 75.50.E; 29.30.H

*Keywords*: Antiferromagnet, Magnetic field, Spin-flop transition, Kinetics of the phase transition, Precessing ball soliton, Evolution of soliton.

## Introduction

The main properties of precessing ball solitons (PBS) in an antiferromagnet at the spin-flop transition, induced by the external magnetic field, have been analyzed in the [1–3] papers. In the given paper, in addition to these papers, we consider the so-called “equilibrium PBS”. Besides, the PBS in the overcritical range of a field, i.e. outside the metastability region, is considered.

## 1. Equilibrium ball solitons

To analyze magnetic solitons in an antiferromagnet, the equations of motion

$$\frac{\partial \mathbf{l}}{\partial t} = \frac{2\mu_B}{\hbar}\left(\mathbf{m}\times\frac{\delta W}{\delta \mathbf{l}} + \mathbf{l}\times\frac{\delta W}{\delta \mathbf{m}}\right) + \Gamma\left(\mathbf{m}\times\frac{\partial \mathbf{l}}{\partial t} + \mathbf{l}\times\frac{\partial \mathbf{m}}{\partial t}\right) \tag{1}$$

$$\frac{\partial \mathbf{m}}{\partial t} = \frac{2\mu_B}{\hbar}\left(\mathbf{m}\times\frac{\delta W}{\delta \mathbf{m}} + \mathbf{l}\times\frac{\delta W}{\delta \mathbf{l}}\right) + \Gamma\left(\mathbf{m}\times\frac{\partial \mathbf{m}}{\partial t} + \mathbf{l}\times\frac{\partial \mathbf{l}}{\partial t}\right). \tag{2}$$

are used [3]. Here W is the thermodynamic potential, **m** and **l** are non-dimensional ferromagnetic and antiferromagnetic vectors; $l_\perp = l_x + il_y, m_\perp = m_x + im_y$ ; the absolute value of the vector **l** at *H=0* equals *1*.

The solutions of (1) and (2) equations in the case of immovable PBS, i.e. at $k \equiv 0$, can be presented in the following form:

$$l_{\perp}(\mathbf{r},\tau) = q(\mathbf{r},\tau)e^{i\omega(\tau)\tau}, \quad m_{\perp}(\mathbf{r},\tau) = p(\mathbf{r},\tau)e^{i\omega(\tau)\tau} \tag{3}$$

and have a spherical symmetry. The corresponding equation has the following form (see [3]):

$$\frac{d^2q}{dr^2} + \frac{2}{r}\frac{dq}{dr} + \frac{q}{1-q^2}\left(\frac{dq}{dr}\right)^2 = q\left(1-q^2\right)\left(1-(\omega+h)^2 - \frac{k_2}{k_1}q^2\right), \tag{4}$$

($k_1$, $k_2$ are the anisotropy constants, $h$ is the external field directed along the anisotropy axis $z$, $h \geq 0$).

The energy of PBS can be expressed as follows:

$$E_s = 8\pi M_0 \alpha_{xy}\sqrt{\frac{\alpha_z}{k_1 B}}\int_0^{\infty}\left\{\left[\frac{1+(\omega+h)^2}{2} - (\omega+h)h\right]q^2 - \frac{k_2}{4k_1}q^4 + \frac{1}{2\left(1-q^2\right)}\left(\frac{dq}{dr}\right)^2\right\}r^2dr, \tag{5}$$

($M_0$ is the magnetization of each sublattice, see other designations in [3]). In Equations (4) and (5), the frequency $\omega$ depends on time.

In the case of immovable PBS the following expressions for the partial are satisfied (see [3]):

$$\frac{\partial q}{\partial \tau} = -\Gamma\omega\left(m_z q + l_z p\right), \tag{6}$$

$$\frac{\partial l_z}{\partial \tau} = 2\Gamma\omega q p, \tag{7}$$

$$\frac{\partial m_z}{\partial \tau} = \Gamma\omega\left(q^2 + p^2\right) \tag{8}$$

$$\frac{\partial p}{\partial \tau} = -\Gamma\omega\left(l_z q + m_z p\right). \tag{9}$$

Follows from these equations that only at the condition $\omega = 0$ all four parameters of PBS are extreme. In this paper we consider the case when this extremum corresponds to the minimum of energy. This is illustrated in Figure 1, where are the frequency dependencies for several values of the field. Indeed, in the field near enough to *1*, the energy of PBS at $\omega = 0$ is minimum. (Here and in all subsequent examples, the parameters, that are typical for antiferromagnetic crystals, have been used, as in [3.4].) This means that at $\omega = 0$ we have an equilibrium state of PBS.

Calculations show that equilibrium PBS are possible only if $h$ value is in a narrow range:

$$1 > h > h^*, \tag{10}$$

where (for our chosen parameters) $h^* = 0.9958$. Maximum amplitude for such PBS equals $q_m = \sqrt{0.5}$ (see [3]).

In Fig.2, the frequency dependencies for several values of the field are shown in the case of reverse phase transition. As can been seen in this Figure, the energy of PBS is minimum at $\omega = 0$ if the field is near enough to $h_{cr} = \sqrt{1 - k_2/k_1}$. In this case, maximum amplitude of equilibrium PBS (at $\omega = 0$) equals $l_{zm} = \sqrt{0.5}$.

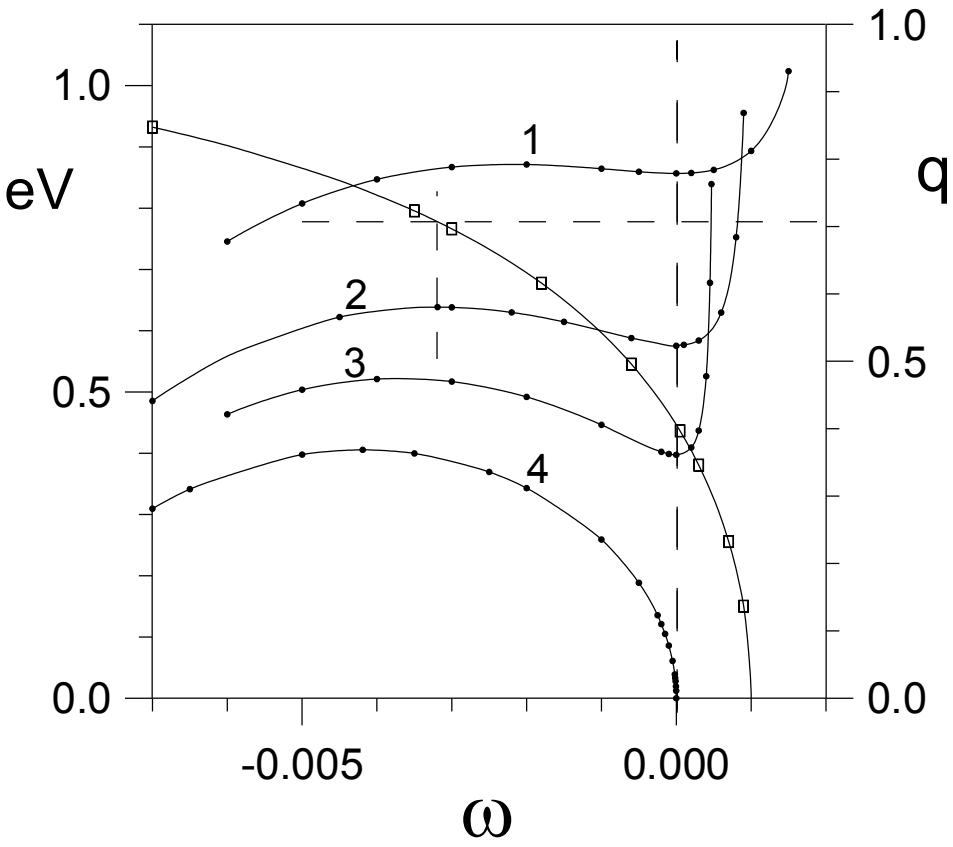

Figure 1: Frequency dependencies of PBS energy for the following values of a field *h*: (1) — 0.998; (2) — 0.999; (3) — 0.9995; (4) — 1. The curve of amplitude is shown by empty squares for $h = 0.999$ only.

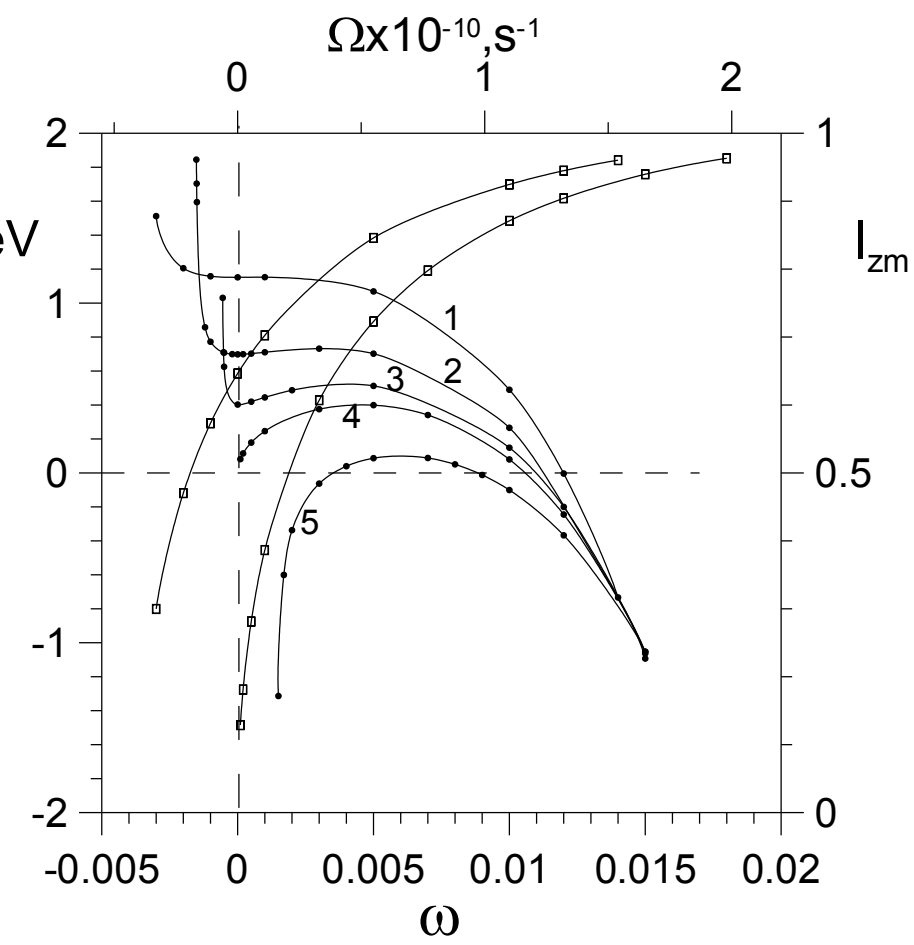

Figure 2: Frequency dependencies of PBS energy for reverse phase transition at the following values of a field *h*: (1) — 0.898; (2) — 0.896; (3) — 0.895; (4) — $h = h_{cr} = \sqrt{1 - k_2/k_1} = 0.894427$, (5) — 0.893. The curves of amplitude are shown for $h = 0.898$ and $h = h_{cr}$.

(It is necessary to note that in Ref. [3] there is an inaccuracy, see Figures 4 and 5 of [3]. The curves of the energy as the function of the frequency can not have the minimum if $h < h^* = 0.9958$.)

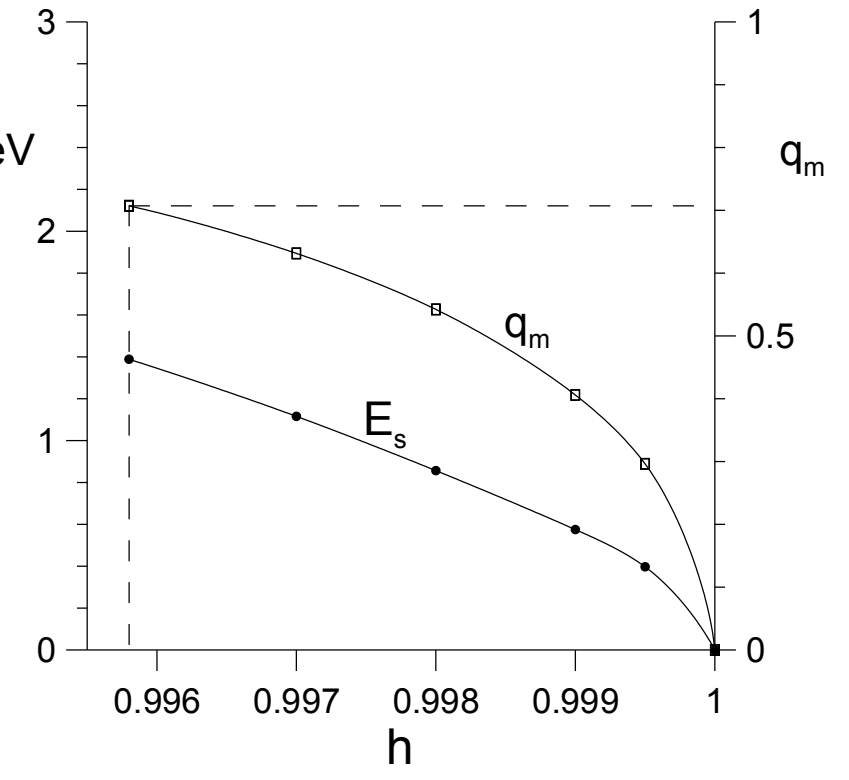

Figure 3: The field dependencies of the energy and amplitude for the equilibrium PBS.

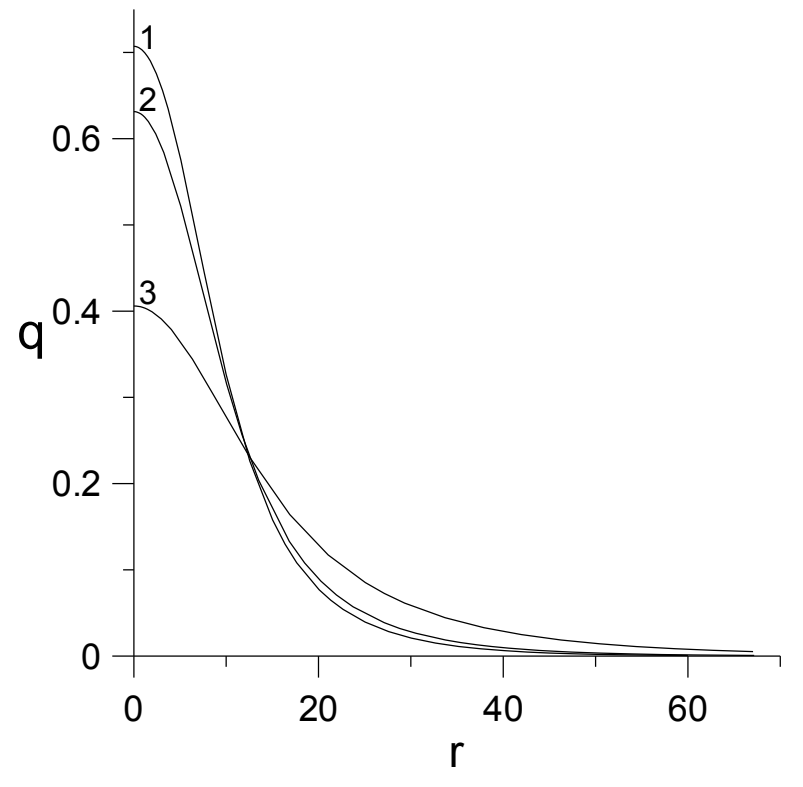

Figure 4: Configurations of the equilibrium PBS: (1) $q_m = \sqrt{0.5}$, (2) $q_m = 0.631$, $h = 0.997$, (3) $q_m = 0.406$, $h = 0.999$.

In Figure 3, the field dependencies of the energy and amplitude for the equilibrium PBS are shown. In Figure 4, the configurations of equilibrium solitons are presented.

Let's assume that at the moment $\tau = 0$ the PBS of the $q(r,0)$ form, corresponding to Equation (4), have arisen. The process of subsequent change of the PBS configuration, in accordance with the following expression:

$$\frac{\partial q}{\partial \tau} \cong -k_1 Q \omega(\omega + h) q(2q^2 - 1) \tag{11}$$

can be obtained using the procedure used in [3]. In Equation (11), the frequency $\omega$ depends on time.

Asymptotical approach of precession frequency to equilibrium value $\omega = 0$ as the time function on the side of lesser and greater values, is shown in Figure 5. Correspondingly, the energy changes according to curve (2) in Figure 1.

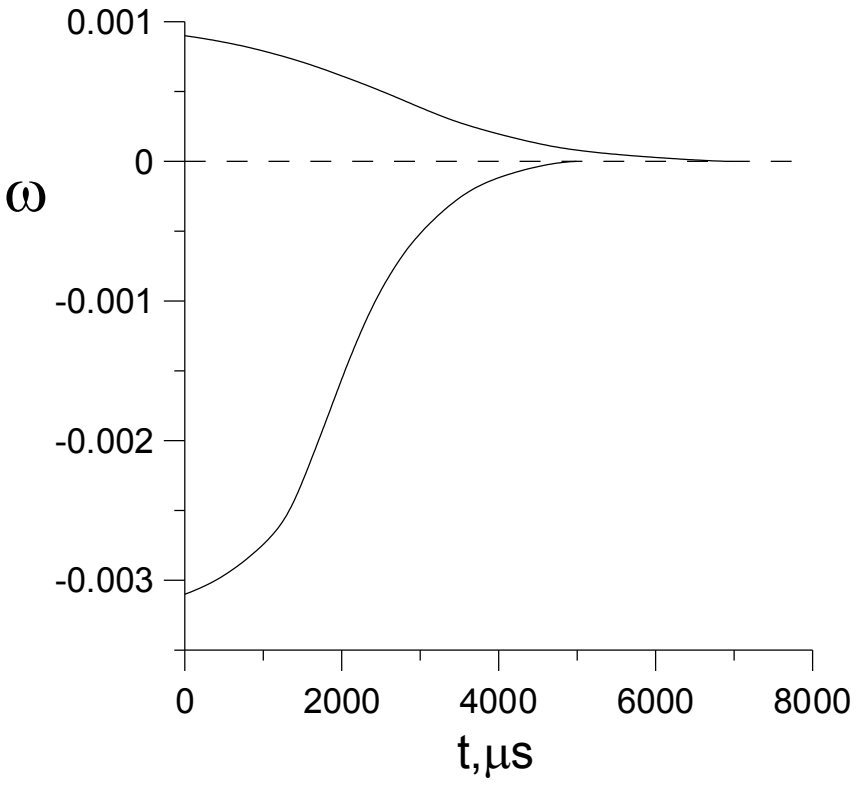

Figure 5: The change of the frequency at approaching of PBS to the equilibrium from the side of lesser and greater values of the frequency if $h = 0.999$.

## 2. Overcritical PBS

The PBS states also exist when the initial-phase state is absolutely unstable [1-3]. Therefore, creation of PBS is possible at disintegration of the initial phase, i.e. at $h > 1$ for the direct phase transition, and at $h < \sqrt{1 - k_2/k_1}$ for reverse spin-flop transition. It is visible in Figure 6 and in curve (5) of Figure 2 for reverse transition. Besides, there is a range of values of a magnetic field where the energy of such PBS can be positive. As seen from Figure 6, for example, at $h = 1.001$ the creation of PBS with near-zero energy is possible for two various frequencies of precession. It can be seen from Equation (4) that for all PBS at $h > 1$ the precession frequencies are negative, but at $h < \sqrt{1 - k_2/k_1}$ for all PBS at reverse transition the precession frequencies are positive.

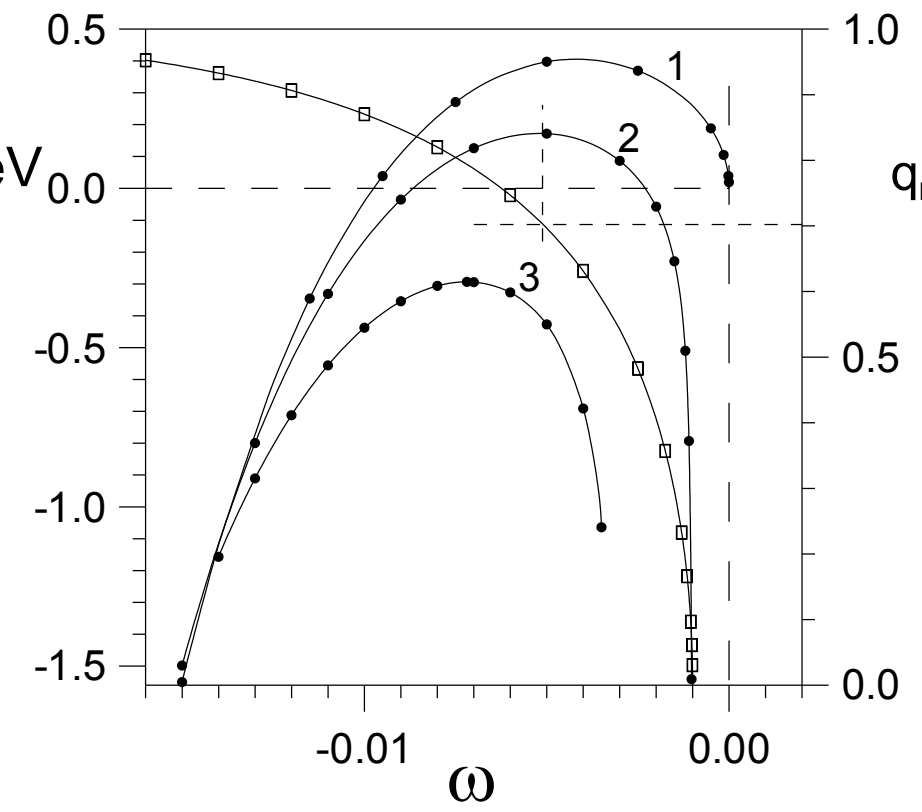

Figure 6: Frequency dependency of PBS energy for the overcritical values of a field: 1 — $h=1$, 2 — $h=1.001$, 3 — $h=1.003$. The curve of amplitude is shown for $h=1.001$. Here and in subsequent figures, the values of energy are denoted by circles; the amplitude, by empty squares; the frequency, by a continuous line; the radius, by crosses.

In Figure 7, the frequency dependencies of energy, amplitude, and radius of PBS at $h=1.001$, $q_m < \sqrt{0.5}$ (it corresponds to curve 2 in Figure 6) are shown. In this case, corresponding to Equation (10), the frequency ω decreases in absolute value, and all $q$ values in PBS, for each radius, decrease, i.e. $\omega \to -0.001$ and $q_m \to 0$. However, the energy gets negative and decreases too. Such change of energy can be explained by the fact that simultaneously the radius of PBS increases considerably. The time dependencies of the main parameters of PBS in this case are presented in Figure 8. Thus, at $h$ value, exceeding the critical value a little, PBS of long duration can originate, which decrease in amplitude and simultaneously increase in radius. Of course, PBS of long duration exist only if they are considered in isolation from other processes. In reality, they will be absorbed by other more quick objects.

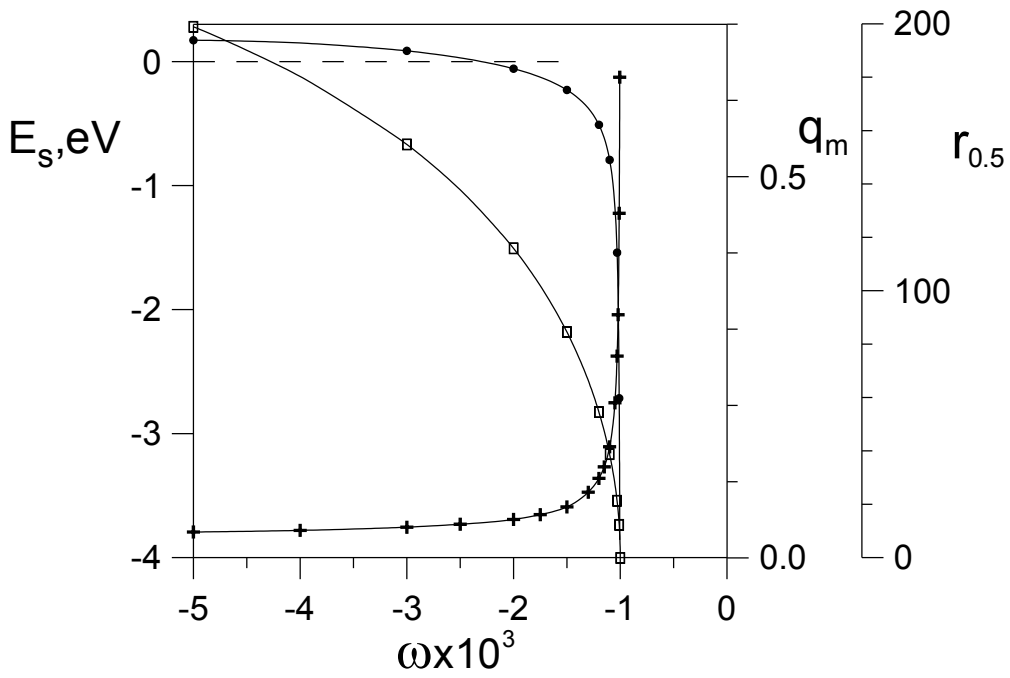

Figure 7: Frequency dependencies of energy, amplitude, and radius of PBS for $h=1.001$, $q_{m,init} < \sqrt{0.5}$. Here the initial amplitude $q_{m,init}=0.696$ (see Figure 6).

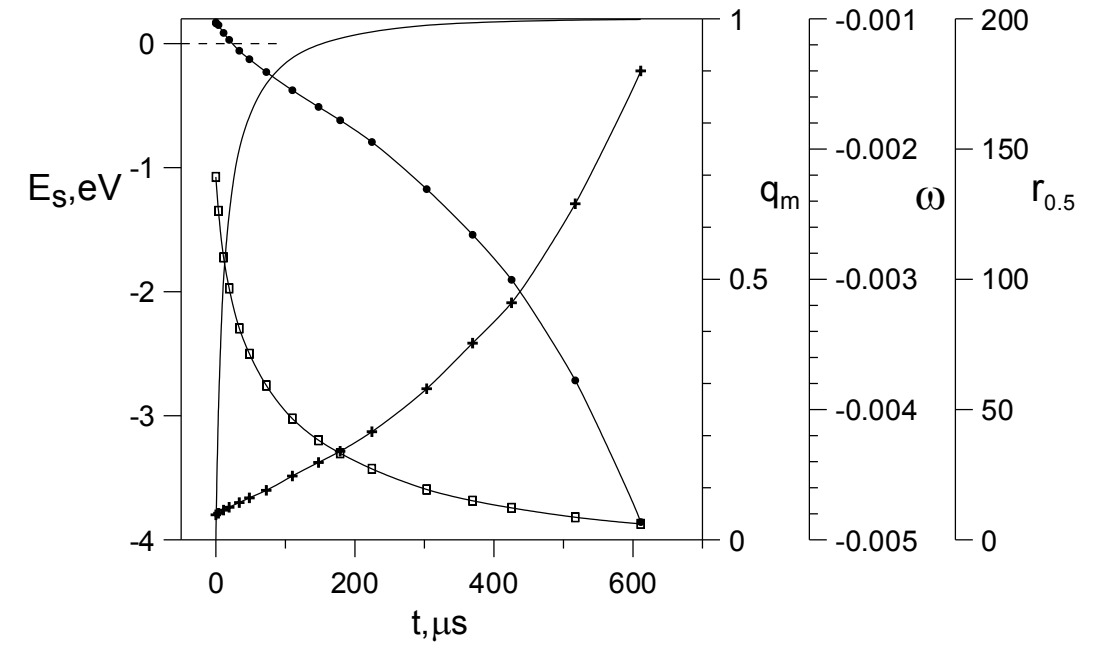

Figure 8: The time dependencies of energy, amplitude, radius, and frequency (continuous line), of PBS if $h=1.001$ and $q_{m,init}=0.696$.

In Figure 9, the time dependencies of energy, amplitude, and precession frequency at $h = 1.001$ , $q_{m,init} > \sqrt{0.5}$ are shown, when the PBS transform "normally" into the domain of the high-field phase.

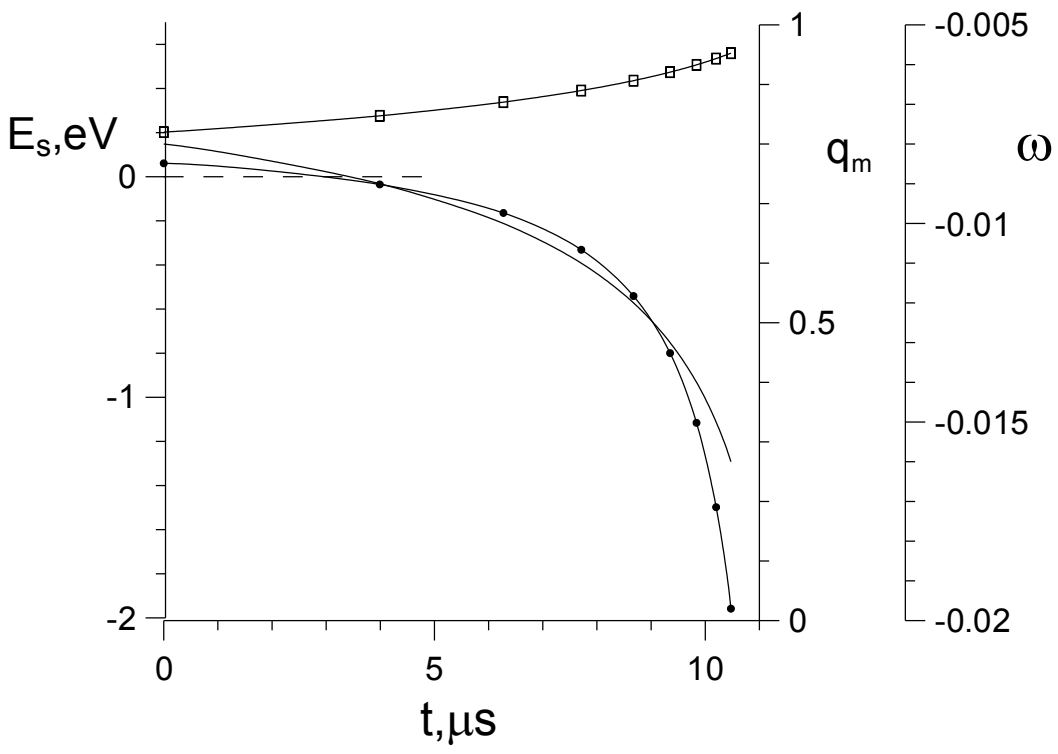

Figure 9: Time dependencies of energy, amplitude, and frequency (continuous line) of PBS if $h = 1.001$ and the initial amplitude $q_{m,init} = 0.82$ (see Figure 6)

## 3. Conclusions

1. At the spin-flop phase transition induced by magnetic field in an antiferromagnet with uni-axial anisotropy, if the field is near enough to critical value, equal to 1 - for the direct phase transition and to $h_{cr} < \sqrt{1 - k_2/k_1}$ - for the reverse spin-flop transition, there is a minimum in the energy of solitons, where the non-precessing equilibrium PBS can exit. The amplitudes of such PBS are limited by value $q_{m\max} \cong \sqrt{0.5}$ for direct spin-flop transition and by $l_{zm\max} \cong \sqrt{0.5}$ - for the reverse transition.
2. In this paper, the process of evolution into such equilibrium PBS from the side of lesser and greater values of the frequency has been analyzed.
3. At $h > 1$ (or $h < \sqrt{1 - k_2/k_1}$ for reverse spin-flop transition), the so-called overcritical PBS are possible. It is supposed that such solitons can originate during the disintegration of the initial phase. For such solitons $\omega < 0$ (or $\omega > 0$ for reverse transition). If amplitude of initial PBS $q_{m,init} > \sqrt{0.5}$ (or $l_{zm,init} > \sqrt{0.5}$ for reverse transition), such solitons transform into domains of the new phase. The relatively small-amplitude solitons, i.e. when the initial amplitude $q_{m,init} < \sqrt{0.5}$ (correspondingly, $l_{zm,init} < \sqrt{0.5}$ for reverse transition), the PBS decrease slowly to zero in amplitude but increase simultaneously in volume. The time dependencies of energy, amplitude, frequency, and radius of such solitons have been presented. It is possible to interpret such solitons originated during disintegration as "softening" of the initial phase